\documentclass[twocolumn,tighten]{aastex631} 

\usepackage{amsmath}

\defcitealias{sato2023exploring}{SK23}

\begin{document}

\title{Spectral Variance in a Stochastic Gravitational-wave Background from a Binary Population}

\author[0000-0003-1096-4156]{William G. Lamb}
\affiliation{Department of Physics \& Astronomy, Vanderbilt University,
2301 Vanderbilt Place, Nashville, TN 37235, USA}
\email{william.g.lamb@vanderbilt.edu}

\author[0000-0001-8217-1599]{Stephen R. Taylor}
\affiliation{Department of Physics \& Astronomy, Vanderbilt University,
2301 Vanderbilt Place, Nashville, TN 37235, USA}
\email{stephen.r.taylor@vanderbilt.edu}

\begin{abstract}
A population of compact object binaries emitting gravitational waves that are not individually resolvable will form a stochastic gravitational wave signal. While the expected spectrum over population realizations is well known from \citet{phinney2001practical}, its higher order moments have not been fully studied before or computed in the case of arbitrary binary evolution. We calculate analytic scaling relationships as a function of gravitational-wave frequency for the statistical variance, skewness, and kurtosis of a stochastic gravitational-wave signal over population realizations due to finite source effects. If the time derivative of the binary orbital frequency can be expressed as a power-law in frequency, we find that these moment quantities also take the form of power-law relationships. We also develop a numerical population synthesis framework against which we compare our analytic results, finding excellent agreement. These new scaling relationships provide physical context to understanding spectral fluctuations in a gravitational-wave background signal and may provide additional information that can aid in explaining the origin of the nanohertz-frequency signal observed by pulsar timing array campaigns.
\end{abstract}

\section{Introduction} \label{sec:intro}
With strong evidence for a nanohertz-frequency gravitational wave (GW) background (GWB) signal recently presented by international collaborations of pulsar timing arrays \citep[PTAs, ][]{ng15-evidence, antoniadis2023second, reardon2023search, xu2023searching}, attention is now shifting to the elusive nature of its origin. While a population of GW-radiating supermassive black-hole binaries (SMBHBs) seems the most likely, many models of early-Universe processes can be formed to explain the PTA data \citep[e.g. ][]{ng15-astro, ng15-newphys, eptadr2-newphys, eptadr2-dm}. 

In the simplest model of an astrophysical GWB one usually assumes a statistically isotropic distribution of SMBHBs across the sky, where the binary orbital evolution at PTA frequencies is driven entirely by the emission of GWs, resulting in a power-law characteristic strain spectrum with $h^2_c(f)\propto f^{-4/3}$ \citep{phinney2001practical}. However this relation is in fact an average over many realizations of an SMBHB population, and several studies have shown that fluctuations in this spectrum are expected \citep[e.g.,][]{sesana2008stochastic,roebber2016cosmic}. 

Recent work has attempted to account for binary population variance in the spectrum and to distinguish between astrophysical and cosmological signal origins. This has included comparing the spectra of cosmological models against a power-law \citep{2022ApJ...938..115K,ng15-newphys}, fitting Gaussian processes or neural networks that are trained on many realizations of simulated astrophysical GWBs to PTA data \citep{ng15-newphys, taylor2017constraints,2024A&A...687A..42B}, searching for anisotropy \citep{ng15-aniso} which may be frequency dependent \citep{2024ApJ...965..164G}, quantifying the discreteness of the GWB at high frequencies \citep{ng15-discrete}, and testing the Gaussian ensemble model given a finite population of GW sources \citep{allen_valtolina_24}.

However, there has been little work to compute the expected variance or higher moments of the GWB signal for a finite population of GW sources. This could be applied to model selection and parameter estimation of the GWB to improve upon the simplistic power-law model for a binary population and provide additional leverage against which other signal models can be compared. There is a brief discussion of this in \citet{2003ApJ...583..616J} in terms of the moment generating function of $h_c^2(f)$, while \citet{satopolito24}\footnote{\citet{satopolito24} use techniques similar to \citet{2020PhRvD.102h3501G} and \citet{2024PhRvD.109h3526G}.} re-analyzed the NANOGrav 15-year data set \citep{ng15-evidence, ng15-astro} with a GWB spectral model that considers deviations from a power-law due to a finite population of SMBHBs. However, explicit scaling relationships for variance, skewness, and kurtosis of the distribution of characteristic strain as a function of frequency were not provided, and the calculations assumed that binaries were only evolving through GW emission in both works. Additionally, \citet{renzini2024projections} investigated the GWB detection capabilities of LIGO-Virgo-KAGRA (LVK) given uncertainties in the source population from single-event analyses. In this letter, we compute the variance and higher-order moments of the GWB for a general finite population of GW sources that create an unresolved GWB. While here we consider only population finiteness effects, we note that GW signal interference will in general also contribute to the variance and higher moments of the GWB. We provide explicit frequency-scaling relationships for arbitrary binary evolution and compare directly to many numerically synthesized populations. These results are relevant not only for SMBHBs in the PTA frequency regime but for other sources of a GWB across the GW frequency spectrum, including LVK \citep[e.g. ][]{abbott2021upper} and LISA \citep{Babak_2023, digman2022lisa, lisa_mission_report}. One can regard this work as a generalization of \citet{phinney2001practical} to higher moments of the GWB strain distribution.

This letter is laid out as follows. In \S\ref{sec:relationships} we introduce a framework to model a finite population of GW sources, which we then use to compute the mean, variance, skewness, and kurtosis of the GWB spectrum across realizations of the population. In \S\ref{sec:pop_synth} we develop a population synthesis model derived from \citet{sato2023exploring} (hereafter \citetalias{sato2023exploring}). The comparison between our analytical and numerical results is shown in \S\ref{sec:results}, followed in \S\ref{sec:discuss} by a discussion of how our new results could aid in answering the origin question for the PTA signal.

\section{GW Strain Spectrum \\from a Binary Population} \label{sec:relationships}
\begin{figure*}
    \centering
    \includegraphics[width=\textwidth]{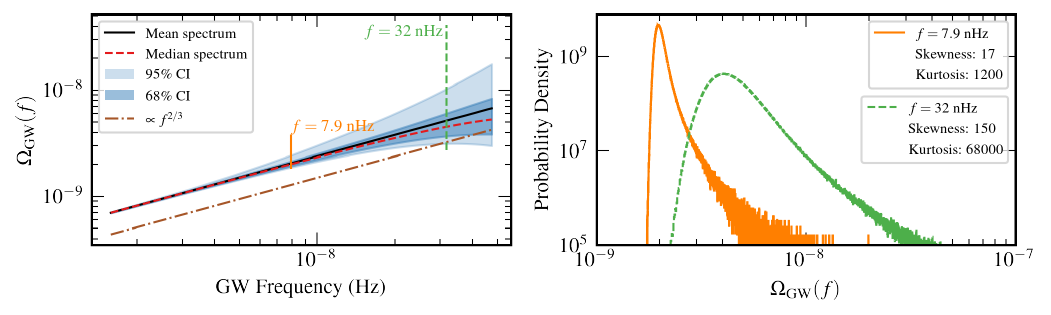}
    \caption{Gravitational wave background from $10^7$ realizations of a binary population described by Model 3 \citetalias{sato2023exploring}. \textit{Left:} We plot the mean and median cosmological energy density as a function of GW frequency. The blue contours represent the 68\% and 95\% confidence intervals of the spectra, and the orange and green lines represent the 99\% confidence interval at two selected frequencies. For comparison, we also plot a $\propto f^{2/3}$ spectrum as expected from a population of circular SMBHB, inspiralling due to GW emission only. \textit{Right:} The GWB distribution across $10^7$ realizations for two selected frequencies, noted in the left plot. As frequency increases, the distribution is wider, hence the variance, skewness, and kurtosis increase.\label{fig:example_spec}
    }
\end{figure*}
The cosmological energy density in GWs, as a fraction of closure density, and as a function of frequency, is given by
\begin{equation} \label{eq:omega}
    \Omega_\mathrm{GW}(f) = \frac{1}{\rho_\mathrm{c}} \frac{\mathrm{d}\rho_\mathrm{GW}}{\mathrm{d}\ln f} = \frac{2\pi^2}{3H_0^2} f^2 h_\mathrm{c}^2(f),
\end{equation}
where $\rho_\mathrm{GW}$ is the energy density of GWs, $\rho_\mathrm{c}$ is the closure density, $f$ is the observed GW frequency, $H_0$ is the Hubble constant, and $h^2_\mathrm{c}(f)$ is the squared characteristic strain. Within a logarithmic frequency bin $\left[ \ln{f} - \case{\Delta\ln{f}}{2}, \ln{f} + \case{\Delta\ln{f}}{2}\right]$, $\Omega_\mathrm{GW}(f)$ is a quadrature sum of GW strain amplitude, $h$, from each source \citep{sesana2008stochastic}:
\begin{align} \label{eq:hc2}
    \Omega_\mathrm{GW}(f) &= \left(\frac{2\pi^2}{3H_0^2}\right)f^2 \int_{\Delta\ln{f}} \frac{\mathrm{d}\ln{f'}}{\Delta\ln{f}} \nonumber\\
    &\times \int_{\vec{\theta}} \mathrm{d}\vec{\theta}' \frac{\mathrm{d}N}{\mathrm{d}\vec{\theta}' \mathrm{d}\ln{f'_r}} h^2(f'_r, \vec{\theta}),
\end{align}
where $N$ is the number of sources, $h$ is the GW strain amplitude from a single GW source, the subscript $r$ denotes the rest-frame of the source such that $f_r=(1+z)f$, and $\vec{\theta}$ are parameters of the source. For example, for populations of SMBHBs, $\vec{\theta}$ includes the chirp mass $\mathcal{M}$ and redshift $z$ of each SMBHB. We denote the binning of sources within a logarithmic frequency bin as an integral over frequency divided by $\Delta \ln{f}$, which effectively averages $\Omega_\mathrm{GW}(f)$ across the log-frequency bin.

We compute the number of binaries per bin of GW frequency and source parameters $\vec{\theta}$ by
\begin{equation}\label{eq:dN}
    \frac{\mathrm{d}N}{\mathrm{d}\vec{\theta}\mathrm{d}\ln{f}} = \frac{\mathrm{d}n}{\mathrm{d}\vec{\theta}}\frac{\mathrm{d}t}{\mathrm{d}\ln{f_r}}\frac{\mathrm{d}z}{\mathrm{d}t}\frac{\mathrm{d}V_c}{\mathrm{d}z},
\end{equation}
where $n$ is the number of SMBHBs per unit comoving volume, $V_\mathrm{c}$. The quantity $\mathrm{d}f/\mathrm{d}t$ describes the dynamical evolution of the SMBHB and is related to the typical amount of time that a binary spends emitting at a given frequency (also known as the residence time), $f/(\mathrm{d}f/\mathrm{d}t)$. For a circular SMBHB evolving entirely due to the emission of GWs, $\mathrm{d}f/\mathrm{d}t \propto f^{11/3}$ \citep{1964PhRv..136.1224P}. Other binary hardening drivers, such as eccentricity of the binary orbit, three-body interactions with stars in the stellar loss cone, or interactions with a circumbinary gas disk, will result in different spectral indices for the power-law relation \citep{kocsis2011gas, sampson2015constraining, burke2019astrophysics}. For example, for three-body stellar scattering, $\mathrm{d}f/\mathrm{d}t \propto f^{1/3}$, while different models of gas interactions can result in $\mathrm{d}f/\mathrm{d}t \propto f^{10/3}$ or $\propto f^{4/3}$. Hence, we model the frequency evolution of the binary as $\mathrm{d}f/\mathrm{d}t \propto f^\lambda$, where $\lambda$ is a free parameter. With $\lambda=11/3$ for GW emission alone, the GWB spectrum has the following power-law form:
\begin{equation}
    \Omega_\mathrm{GW}(f) = \frac{2\pi^2}{3H_0^2}A^2_\mathrm{ref}f^2\left(\frac{f}{f_\mathrm{ref}}\right)^{2\alpha},
\end{equation}
where $A_\mathrm{ref}$ is the strain amplitude measured at a reference frequency $f_\mathrm{ref}$, and $\alpha=(7/3-\lambda)/2=-2/3$.

The original calculation from \citet{phinney2001practical} can be further generalized by considering the distribution of SMBHB source positions,
\begin{align} \label{eq:harmonics}
    \Omega_\mathrm{GW}(f, \vec{\theta}) &= \int_{\hat{\Omega}} \Omega_\mathrm{GW}(f, \vec{\theta}, \hat{\Omega}') \mathrm{d}\hat{\Omega}' \nonumber \\
    &= \frac{2\pi^2}{3H_0^2}f^2\int_{\Delta\ln{f}} \frac{\mathrm{d}\ln{f}'}{\Delta\ln{f}} \int_{\vec{\theta}} \mathrm{d}\vec{\theta}' \nonumber \\
    &\ \ \ \ \times \int_{\hat{\Omega}} \frac{\mathrm{d}\hat{\Omega}'}{4\pi} \frac{\mathrm{d}N}{\mathrm{d}\vec{\theta}'\mathrm{d}\ln{f_r}'\mathrm{d}\hat{\Omega}'} h^2\left(f_r', \vec{\theta}'\right),
\end{align}
where $\hat\Omega$ is the direction of propagation of a GW. 

\citetalias{sato2023exploring} adopt this form and discretize the GW source populations as follows. The sky is divided up into infinitessimal cells with unit vector $\hat{\Omega}_i$ pointing towards the center of cell $i$. In each cell, there is exactly zero or one GW source within the frequency bin $\left[\ln{f} - \case{\Delta\ln{f}}{2}, \ln{f} + \case{\Delta\ln{f}}{2}\right]$. If $N_i$ is the number of binaries in cell $i$, then the expected number of binaries in each cell is
\begin{equation} \label{eq:expect_N}
    \left<N_i(f_i, \vec{\theta}_i)\right>\mathrm{d}\vec{\theta}_i = \frac{1}{4\pi \Delta\ln{f_i}} \int_{\Delta\ln{f_i}} \mathrm{d}\ln{f_i}' \frac{\mathrm{d}N}{\mathrm{d}\vec{\theta}_i'\mathrm{d}\ln{f_{r,i}'}} \mathrm{d}\vec{\theta}_i',
\end{equation}
where $\left<\cdot\right>$ denotes expectation over sky position. Note that, by definition, $N_i^2=N_i$. Cells $i$ and $j$ should be uncorrelated, thus for a frequency bin $\left[\ln{f} - \case{\Delta\ln{f}}{2}, \ln{f} + \case{\Delta\ln{f}}{2}\right]$:
\begin{equation}\label{eq:prop_N}
    \left<N_i N_j\right> = \begin{cases}
        \left<N_i\right>\left<N_j\right> &\text{for $i\neq j$} \\
        \left<N_i^2\right>\delta_{\vec{\theta}_i \vec{\theta}_j} = \left<N_i\right>\delta_{\vec{\theta}_i \vec{\theta}_j} &\text{for $i=j$},
    \end{cases}
\end{equation}
where $\delta$ is the Kronecker delta. For a cell $i$, the expected energy density of the GWB is $\left<\Omega_\mathrm{GW}(f_i, \vec{\theta}_i, \hat{\Omega}_i)\right> \propto f^2 \left<h^2_\mathrm{c}(f_i, \vec{\theta}_i, \hat{\Omega}_i)\right> = \left<N_i(f_i, \vec{\theta}_i)\right> h^2(f_i, \vec{\theta}_i)$, and the expected energy density across the sky is given by
\begin{align} \label{eq:expected_hc2}
    \left<\Omega_\mathrm{GW}(f_i)\right> &= \frac{2\pi^2}{3H_0^2}f^2 \int \mathrm{d}
    \hat{\Omega}_i' \int_{\vec{\theta}_i} \mathrm{d}\vec{\theta}_i' \left<N_i(f_i, \vec{\theta}_i')\right> h^2(f_i, \vec{\theta}_i') \nonumber \\
    &= \frac{2\pi^2}{3H_0^2}f^2 \int_{\vec{\theta}_i} \mathrm{d}\vec{\theta}_i' \left<N(f_i, \vec{\theta}_i')\right> h^2(f_i, \vec{\theta}_i'),
\end{align}
where $\left<N\right>=\int\mathrm{d}\hat{\Omega}_i\left<N_i\right> = 4\pi\left<N_i\right>$, the expected number of binaries in a given realization. This agrees with \autoref{eq:hc2}.
For the second non-central moment of $h_\mathrm{c}^2(f)$,
\begin{align} \label{eq:corr_hc2}
    &\left<h^2_\mathrm{c}(f_i, \vec{\theta}_i, \hat{\Omega}_i) h^2_\mathrm{c}(f_j, \vec{\theta}_j, \hat{\Omega}_j)\right> \nonumber\\
    &= \delta_{f_i f_j} \left<N_i(f_i, \vec{\theta}_i)N_j(f_j, \vec{\theta}_j)\right> h^2(f_i, \vec{\theta}_i)h^2(f_j, \vec{\theta}_j) \nonumber \\
    &= \delta_{f_i f_j}\left<N_i(f_i, \vec{\theta}_i)\right>\left<N_j(f_j, \vec{\theta}_j)\right> h^2(f_i, \vec{\theta}_i) h^2(f_j, \vec{\theta}_j) \nonumber \\
    &\ \ \ \ + \delta_{ij}\delta_{f_i f_j}\delta_{\vec{\theta}_i\vec{\theta}_j}\left<N_i(f_i, \vec{\theta}_i)\right> h^4(f_i, \vec{\theta}_i).
\end{align}
Integrating over sky-positions $\hat{\Omega}_i$ and $\hat{\Omega}_j$ and binary parameters $\vec{\theta}_i, \vec{\theta}_j$, and substituting \autoref{eq:prop_N} gives,
\begin{align} \label{eq:centralmoment2}
    \left<h^2_\mathrm{c}(f_i)h^2_\mathrm{c}(f_j)\right> &= \delta_{f_i f_j}\left( \int_{\vec{\theta}} \mathrm{d}\vec{\theta}'\left<N(f_i, \vec{\theta}')\right> h^2(f_i,\vec{\theta}')\right)^2 \nonumber \\
    &+ \delta_{f_i f_j} \int_{\vec{\theta}} \mathrm{d}\vec{\theta}'\left<N(f_i, \vec{\theta}')\right> h^4(f_i, \vec{\theta}')
\end{align}
Integrating over frequency bin $\Delta\ln{f_j}$, we can then compute the variance on $\Omega_\mathrm{GW}(f)$:
\begin{align} \label{eq:var_hc2}
    \mathrm{Var}\left[\Omega_\mathrm{GW}(f)\right] &= \left(\frac{2\pi^2f^2}{3H_0^2}\right)^2 \left[\left<\left(h_\mathrm{c}^2(f)\right)^2\right> - \left<h_\mathrm{c}^2(f)\right>^2\right] \nonumber \\
    &= \frac{4\pi^4}{9H_0^4}\frac{f^4}{\Delta\ln{f}}\int \mathrm{d}\vec{\theta}\left<N(f, \vec{\theta})\right> h^4(f, \vec{\theta}) \nonumber\\
\end{align}
Since $\left<N\right> \propto \mathrm{d}N / \mathrm{d}\ln{f} \propto f \mathrm{d}t/\mathrm{d}f_r \propto f^{1-\lambda}$, $\Delta\ln{f} \propto 1/f$, and $h(f) \propto f^{2/3}$, we get the following relation:
\begin{equation}
    \mathrm{Var}\left[\Omega_\mathrm{GW}(f)\right] \propto f^4 \cdot f \cdot f^{1-\lambda} \cdot f^\frac{8}{3} \propto f^{\frac{26}{3}-\lambda}.
\end{equation}

We derive the mean and variance of various useful parameters related to $\Omega_\mathrm{GW}(f)$ in \autoref{tab:rel_table}, using the standard transformation $\mathrm{Var}\left[aX+b\right] = a^2\mathrm{Var}\left[X\right]$ for random variable $X$ and constants $a, b$. Note that our results agree with the frequency relation in Equation 29 of \citet{mingarelli2013characterizing}, where they compute $\sigma\left(\Omega_\mathrm{GW}(f)\right)/\Omega_\mathrm{GW}(f) \propto f^{1/2}/f^{-4/3} = f^{11/6}$, where $\sigma^2=\mathrm{Var}$. We also see that upon taking the ratio of the variance to the mean, the frequency scaling no longer depends on the binary evolution parameter $\lambda$, thus creating a clean probe of the Poissonian statistics of the binary population without the need to specify the dynamical mechanisms of binary decay.

The calculations for skewness and excess kurtosis on $\Omega_\mathrm{GW}(f)$ proceed in much the same way. The skewness is a measure of the asymmetry of a distribution about its mean, where a positive skewness indicates that most of the mass of the distribution is to the right of the mean, and \textit{vice versa}. The excess kurtosis is a measure of the extremities of deviations from the mean (i.e. outliers) relative to a Gaussian distribution, where a positive kurtosis indicates more extreme outliers resulting in wider tails, and a negative kurtosis indicates fewer extreme outliers and a thinner tail, than a Gaussian distribution. Note that $\mathrm{Skew}[aX+b] = \mathrm{Skew}[X]$ and $\mathrm{Kurt}[aX+b] = \mathrm{Kurt}[X]$, therefore the skewness and kurtosis are equal between linear transforms of $\Omega_\mathrm{GW}$.
\begin{gather}
    \mathrm{Skew}(f) = \left(\frac{1}{\Delta\ln{f}}\right)^{\frac{1}{2}}\frac{\int \mathrm{d}\vec{\theta}\left<N(f, \vec{\theta})\right> h^6(f, \vec{\theta})}{\left(\int \mathrm{d}\vec{\theta}\left<N(f, \vec{\theta})\right> h^4(f, \vec{\theta})\right)^{\frac{3}{2}}} \label{eq:skew}\\
    \mathrm{Kurt}(f) = \frac{1}{\Delta\ln{f}}\frac{\int \mathrm{d}\vec{\theta}\left<N(f, \vec{\theta})\right> h^8(f, \vec{\theta})}{\left(\int \mathrm{d}\vec{\theta}\left<N(f, \vec{\theta})\right> h^4(f, \vec{\theta})\right)^2}\label{eq:kurt}
\end{gather}
The skewness and kurtosis are always positive. We derive the following frequency relations which are relevant for all linear transformations of $\Omega_\mathrm{GW}$:
\begin{gather}
    \mathrm{Skew}(f) \propto f^\frac{1}{2} \cdot f^{1-\lambda} \cdot f^{4} \cdot (f^{1-\lambda} \cdot f^\frac{8}{3})^{-\frac{3}{2}} \propto f^{\frac{\lambda}{2}} \\
    \mathrm{Kurt}(f) \propto f \cdot f^{1-\lambda} \cdot f^\frac{16}{3} \cdot (f^{1-\lambda} \cdot f^\frac{8}{3})^{-2} \propto f^{\lambda}.
\end{gather}
We now see that taking the ratio between the kurtosis and the squared skewness removes all frequency dependence, providing a clear prediction for a stochastic GW signal generated by a Poisson-distributed binary population.
\begin{deluxetable}{C|C|C|C}
    \tablecaption{A table of relationships between different parameters relative to the squared characteristic strain, $h_c^2$, including fractional energy density, $\Omega_\mathrm{GW}$, one-sided power spectral density of the GWB Fourier modes, $S_h$, and one-sided power spectral density of pulsar timing residuals induced by a GWB, $S_{\delta t}$. Assumes an SMBHB frequency evolution of $\mathrm{d}f/\mathrm{d}t \propto f^\lambda$. Skewness and kurtosis scale as $f^{\lambda/2}$ and $f^\lambda$, respectively, for all quantities. \label{tab:rel_table}}
    \tablecolumns{4}
    \tablehead{
         \colhead{} & \colhead{Scaling} & \colhead{Mean} & \colhead{Variance}
    }
    \startdata
        $\Omega_\mathrm{GW}(f)$ & - & f^{13/3 - \lambda} & f^{26/3 - \lambda} \\ 
        h_\mathrm{c}^2(f) & f^{-2}\Omega_\mathrm{GW} & f^{7/3 - \lambda} & f^{14/3 - \lambda} \\ 
        S_h(f) & f^{-3}\Omega_\mathrm{GW} & f^{4/3 - \lambda} & f^{8/3 - \lambda} \\
        S_{\delta t}(f) & f^{-5}\Omega_\mathrm{GW} & f^{-2/3 - \lambda} & f^{- 4/3 - \lambda}
    \enddata
\end{deluxetable}
\section{Population Synthesis}\label{sec:pop_synth}
We simulate $10^7$ realizations of a population of SMBHBs. We use a Schechter-based model to describe the number density of SMBHBs per unit redshift and chirp mass \citep{middleton16} \citep[see also][]{sesana2008stochastic, sato2023exploring, ng15-astro}:
\begin{align}\label{eq:middleton}
    \frac{d^2n}{dzd\log\mathcal{M}} = &\dot{n}_0\left[\left(\frac{\mathcal{M}}{10^7M_\odot}\right)^{-\alpha}e^{-\mathcal{M}/\mathcal{M}_*}\right] \nonumber \\
    &\times \left[\left(1+z\right)^\beta e^{-z/z_0}\right]\frac{dt_r}{dz},
\end{align}
where $\dot{n}_0$ is the normalised merger rate, $\mathcal{M}_*$ is the characteristic mass, $\alpha$ is the mass spectral index, $z_0$ is the characteristic redshift, and $\beta$ is the redshift spectral index.  (See \citet{middleton16} for details). To create the population of SMBHBs, we Poisson sample the number of sources in each frequency, chirp mass, and redshift bin such that \citep{sesana2008stochastic, middleton16, ng15-astro}
\begin{align}
    h_c^2(f) = \sum_{\log\mathcal{M}, z} &\mathcal{P}\left(\frac{\mathrm{d}^3N}{\mathrm{d}\log{\mathcal{M}}\mathrm{d}z\mathrm{d}\ln{f_r}}\Delta\log{\mathcal{M}}\Delta z \Delta\ln{f}\right) \nonumber \\
    &\times \frac{h^2(f_r, \log\mathcal{M}, z)}{\Delta\ln{f}}.
\end{align}

In our analyses, we implement the five different models used in \citetalias{sato2023exploring}, calibrating $\dot{n}_0$ such that $A_{1\mathrm{ yr}^{-1}} \approx 2\times10^{-15}$ at $f=1\mathrm{ yr}^{-1}$, as was found in \citet{ng15-evidence}. We model a circular binary whose evolution is purely driven by GW emission, hence $\lambda = 11/3$. Our population synthesis code is available on \href{https://github.com/astrolamb/pop_synth}{\texttt{Github}}.

\section{Results}\label{sec:results}
\begin{figure*}
    \centering
    \includegraphics[width=\textwidth]{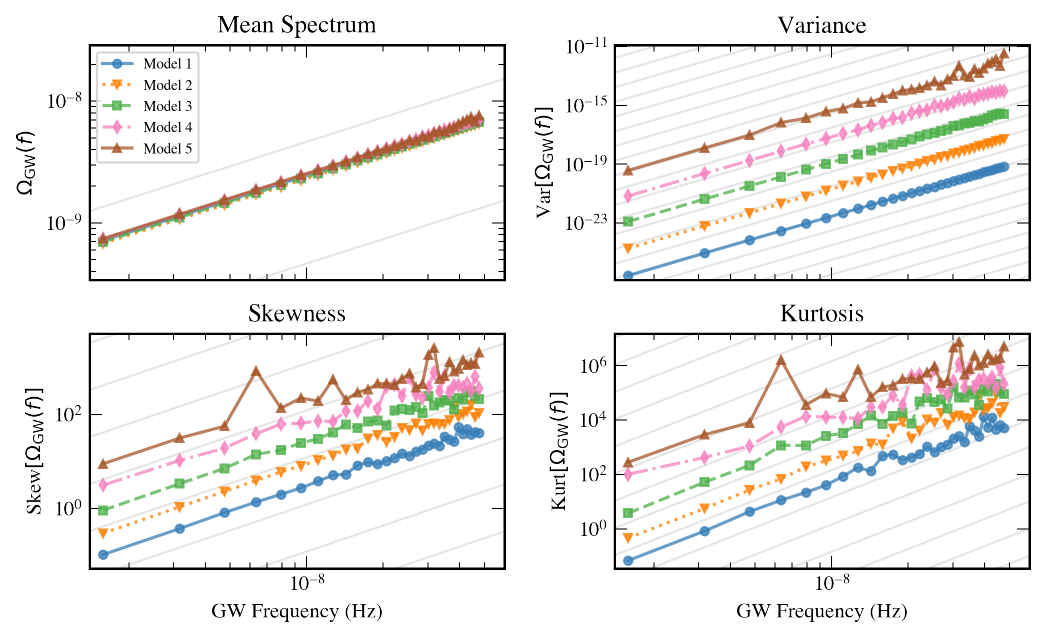}
    \caption{Comparing the mean, variance, skewness, and kurtosis of $10^7$ realizations of the GWB as a function of GW frequency for the five models defined in \citetalias{sato2023exploring}. The grey lines represent the general frequency scaling relation for the given statistic in the case of $\lambda=11/3$, which are $\Omega_\mathrm{GW}(f) \propto f^{2/3}$, $\mathrm{Var[\Omega_\mathrm{GW}]}\propto f^5$, $\mathrm{Skew} \propto f^{11/6}$, and $\mathrm{Kurt} \propto f^{11/3}$. For all models, we see strong agreement with the relations found in \autoref{tab:rel_table}.\label{fig:stats}}
\end{figure*}
We now compare the $10^7$ realizations of synthesized SMBHB populations against our analytical results. Given that $\lambda=11/3$ for circular SMBHBs hardened by GW emission, from \autoref{tab:rel_table} we expect the following frequency relationships in our results: $\Omega_\mathrm{GW}(f) \propto f^{2/3}$, variance $\mathrm{Var}\left[\Omega_\mathrm{GW}(f)\right] \propto f^5$; $\mathrm{Skew}(f) \propto f^{11/6}$; $\mathrm{Kurt}(f) \propto f^{11/3}$.

In \autoref{fig:example_spec}, we show the mean, median, $68\%$ and $95\%$ confidence intervals of simulated GWB spectra across $10^7$ realizations of Model 3 from \citetalias{sato2023exploring} in terms of $\Omega_\mathrm{GW}(f)$. We see that the mean follows the expected $\Omega_\mathrm{GW}(f) \propto f^{2/3}$ relation that we would expect from a GWB sourced from a finite population of circular SMBHBs, while the spread in the spectrum increases as frequency increases. We show the 99\% confidence interval for two frequencies, $f=7.9$ nHz and $f=32$ nHz, as vertical lines to further show the increase in the width of the distribution as frequency increases. In the right-hand panel, we show the shape of the distribution of $\Omega_\mathrm{GW}(f)$ across the $10^7$ realizations at these two frequencies. We see that as frequency increases, the distribution widens, hence the variance increases. The mass of the distributions is also to the left of the mean, and they become more asymmetrical and feature more extreme outliers (as shown by the wider tail) as frequency increases. This suggests positive skewness and positive excess kurtosis which increases with increasing frequency, which agrees with our interpretation of \autoref{eq:skew} and \autoref{eq:kurt}.

\begin{figure*}
    \centering
    \includegraphics[width=\textwidth]{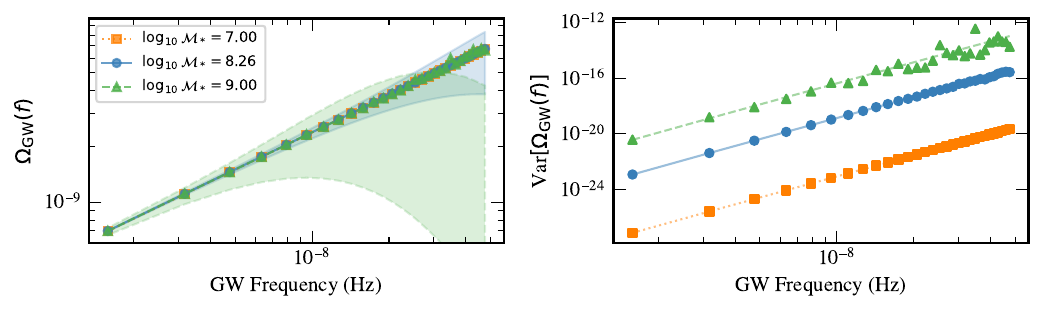}
    \caption{The characteristic mass $\mathcal{M}_*$ has a strong effect on the variance of the GWB. We use `Model 3' of \citetalias{sato2023exploring}, but change $\mathcal{M}_*$ (and $\dot{n}_0$). \textit{Left}: The mean $\Omega_\mathrm{GW}(f)$ as a function of frequency for the three submodels, along with the $68\%$ confidence interval of the GWB (shaded contours). As $\mathcal{M}_*$ increases, the width of the $68\%$ C.I. increases. Note that the mean spectrum for $\log\mathcal{M}_*=9.00$ is outside of the 68\% C.I. at high frequencies. \textit{Right:} The variance of the GWB as a function of frequency for the three submodels. The markers are the measured variance from $10^7$ realizations. The plotted lines are the analytically computed variance, given by \autoref{eq:var_hc2}, for each model. The measured variance and analytical variance are strongly consistent.\label{fig:char_mass}}
\end{figure*}
\autoref{fig:stats} shows the mean, variance, skewness, and kurtosis of our $10^7$ GWB realizations as a function of frequency for the five models defined in \citetalias{sato2023exploring}. Since we defined $\dot{n}_0$ in each model such that $A_{1 \mathrm{ yr}^{-1}} \approx 2\times10^{-15}$ \citep{ng15-evidence}, it is expected, and observed, that the mean spectrum of each model is identical. Comparing the computed variance of the GWB to the guiding lines of $\propto f^5$ shows excellent consistency. As we get to higher-order moments, the effect of having a finite number of samples to approximate and analyze the distribution becomes apparent. For Models 1 and 2, the skewness and kurtosis follow the expected relation very well, despite some instability at higher frequencies. The skewness of Model 3 follows the frequency relation well across all frequencies too, however, for high frequencies, the kurtosis slightly deviates from the expected $f^{11/6}$ relation. For models 4 and 5, the skewness and kurtosis follow the correct frequency relation at lower frequencies but deviate from this relation at higher frequencies. We attribute this simply to a limited sample size. Some of these models have very long tails dominated by rare massive binaries, and sampling a population realization where the GWB at this frequency deviates so far from the mass of the distribution to explore these tails is extremely unlikely. Hence, we underestimate the skewness and kurtosis at these frequencies. To create $10^7$ realizations, we ran our code for 12 hours across 100 CPUs for each model. Running at least $10^{8-9}$ realizations could begin to ameliorate this, however, this would be computationally expensive. We also note that in Model 5 at $\sim6$~nHz there is a deviation from the expected skewness and kurtosis; this is due to a single extreme outlier in one realization. Removing this point removes this deviation, and the measured skewness and kurtosis are consistent with our expectations. 

The magnitude (and numerical stability) of the variance, skewness, and kurtosis at high frequencies are model dependent. The populations that result in the largest variance and instability in the spectra are models that have a larger characteristic mass $\mathcal{M}_*$ which controls the exponential cutoff to the number of high-mass binaries in \autoref{eq:middleton}. In these models where very high mass SMBHBs are more likely, their large strain amplitudes cause them to dominate the GWB spectrum such that the expected number of SMBHBs in these frequency bins is $\mathcal{O}(1)$. Therefore, some realizations will not have a binary in these bins at all, while others will, resulting in a large variance (and skewness and kurtosis) of the GWB spectrum. To investigate the effect of the characteristic mass on the distribution of the spectrum, we create two sets of $10^7$ realizations of a Model 3 population, where we change the characteristic mass to $\log\mathcal{M}_*=7$ and $\log\mathcal{M}_*=9$ (and normalize $\dot{n}_0$ as required to maintain $A_{1\mathrm{ yr}^{-1}}$).

We show the results of these simulations in \autoref{fig:char_mass}. Increasing the characteristic mass increases the variance of the GWB, which strongly follows the $f^5$ dependence from \autoref{tab:rel_table}. For the model with $\log\mathcal{M}_*=9$, we see in the left panel that the mean spectrum across realizations is outside of the 68\% confidence interval, suggesting strong influence by a small number of realizations with large $\Omega_\mathrm{GW}$ at high frequencies, corresponding to a small number of SMBHBs in those realizations. However, for the majority of realizations there are no binaries and hence no GWB in these frequency bins, causing the median spectrum to depart from the mean. 

Decreasing $\log\mathcal{M}_*$ results in decreasing variance across the spectrum because of the exponential cut-off at a relatively low mass, which leads to a large number of low mass (hence low strain amplitude) binaries to create a GWB consistent with $A_{1\mathrm{ yr}^{-1}}=2\times10^{-15}$. The 68\% confidence interval for $\log\mathcal{M}_*=7$ is very narrow and cannot be seen in \autoref{fig:char_mass}. In the right panel of \autoref{fig:char_mass}, we compare the measured variance for the three choices of $\log\mathcal{M}_*$ (shown by markers) against the analytically derived variance in \autoref{eq:var_hc2} (plotted lines), showing excellent consistency. 

\section{Discussion}\label{sec:discuss}
We derived analytical scaling relationships for the variance, skewness, and kurtosis of the GWB spectrum across realizations of a finite population of GW sources for the first time. For a binary population hardened by GW emission, we find that the mean, variance, skewness, and kurtosis of $\Omega_\mathrm{GW}(f)$ scale as $f^{2/3}$, $f^5$, $f^{11/6}$, and $f^{11/3}$, respectively.  
We created simulated populations of SMBHBs and compared the spectral moments to our analytical forms, finding excellent agreement.

In addition, our approach can compute the moment statistics of the GWB spectrum for any $\mathrm{d}f/\mathrm{d}t$. For example, we expect that SMBHBs interact with their galaxy environments at larger separations (hence lower GW frequencies), resulting in accelerated GW evolution in this frequency range. Assuming that each mechanism being considered follows a power-law relation $f^{\lambda_i}$ for each mechanism $i$, we can model $\mathrm{d}f/\mathrm{d}t$ as 
\begin{equation} \label{eq:dfdt_sum}
    \frac{\mathrm{d}f}{\mathrm{d}t} = g(f|\{c_i,\lambda_i\}) = \sum_i c_i f^{\lambda_i},
\end{equation}
for coefficients $c_i$, which depend on other binary and environmental properties. Furthermore, while this letter considers only circular binary populations, one could extend our formalism to populations with eccentric orbits.

We have shown our analytical results for linear transformations of the GW cosmological energy density, $\Omega_\mathrm{GW}(f)$. However, other useful quantities involve non-linear transformations of $\Omega_\mathrm{GW}(f)$, the most notable of which is the RMS characteristic strain $h_\mathrm{c}(f)$. Transforming these variables requires an expansion of the variable about the mean, and applying the statistic to that expansion. To leading order this gives $\mathrm{Var}[h_c(f)]\propto f^{7/3}$ for GW-hardened binaries, but this assumes that fluctuations about the mean are small. We have tested this against our numerical simulations, observing good agreement at low frequencies but poor performance at higher frequencies for models in which large strain outliers are more probable, e.g., low characteristic redshift and high characteristic chirp mass.

These results could be used in exploring the origin of GWB signals. For example, in PTA science the standard astrophysical GWB model is a power-law spectrum. However, a single population realization's spectrum will deviate from a power-law. Hence, including information on higher-order moment statistics could lead to more robust analyses to account for spectral excursions. One could directly fit SMBHB population models to PTA data with this information \citep{satopolito24}, bypassing computationally expensive simulations and machine learning \citep{ng15-astro, taylor2017constraints, shih2023fast} using the methods developed in \citet{lamb2023rapid}. This will enable more refined GWB and noise estimation in PTA data if the GWB is assumed to be of astrophysical origin. 
Indeed by adopting a more general dynamical model like \autoref{eq:dfdt_sum} in our calculations, one could leverage higher statistical moment information to probe the dynamics of SMBHB evolution in the sub-parsec regime.

PTA spectral characterization is likely too insensitive at present to accurately measure skewness and kurtosis, but these higher-order moments could be useful in the future. Meanwhile, the future LISA mission will be sensitive to several stochastic signals, such as cosmological stochastic backgrounds from first-order phase transitions, cosmic strings, and primordial black holes, a stochastic foreground from Galactic white-dwarf binaries, and potentially backgrounds from extragalactic stellar-mass black-hole binaries and extragalactic white-dwarf binaries \citep{Babak_2023, farmer2003gravitational, lisa_mission_report}. Incorporating information on the statistics of spectral deviations from simple power-law models could help LISA analyses distinguish astrophysical from cosmological GWB signals.

Finally, beyond the population finiteness effects studied here, the additional influence of GW signal interference between sources will contribute as a source of (cosmic) variance \citep{allen_valtolina_24,2023PhRvD.107d3018A} and potentially to higher statistical moments of the GWB spectrum. Frequency-scaling relationships for signal interference effects, and the relative importance of population finiteness and signal interference, will be explored in future work.

\begin{acknowledgements}
    The authors wish to thank Kai Schmitz for fruitful discussions about this work. The authors are members of the NANOGrav collaboration, which receives support from NSF Physics Frontiers Center award number 1430284 and 2020265. SRT acknowledges support from an NSF CAREER \#2146016, NSF AST-2007993, and NSF AST-2307719. WGL acknowledges support from the Vanderbilt Physics \& Astronomy McMinn Summer Research Award. This work was conducted in part using the resources of the Advanced Computing Center for Research and Education (ACCRE) at Vanderbilt University, Nashville, TN.
\end{acknowledgements}
\software{numpy \citep{numpy}, matplotlib \citep{plt}}
\bibliography{sample631}{}
\bibliographystyle{aasjournal}
\end{document}